\documentclass[superscriptaddress,twocolumn,aps,amsmath,amssymb]{revtex4}
\usepackage{graphicx,color}
\usepackage{bm}
\usepackage{hyperref}

\newcommand{\Sec}[1]{Sec.\,\ref{#1}}
\newcommand{\App}[1]{Appendix\,\ref{#1}}

\newcommand{\be}{\begin{equation}}
\newcommand{\ee}{\end{equation}}
\newcommand{\bsube}{\begin{subequations}}
\newcommand{\esube}{\end{subequations}}

\newcommand{\Eq}[1]{Eq.\,(\ref{#1})}
\newcommand{\Eqs}[1]{Eqs.\,(\ref{#1})}

\newcommand{\dg}{\dagger}
\newcommand{\la}{\langle}
\newcommand{\ra}{\rangle}

\newcommand{\ti}{\Tilde}
\newcommand{\nl}{\nonumber \\}
\newcommand{\nla}{\nl&\quad}
\newcommand{\up}{\uparrow}
\newcommand{\down}{\downarrow}

\newcommand{\leftact}{\overset{\rightarrow}}
\newcommand{\rightact}{\overset{\leftarrow}}

\begin{document}
\draft

\title{Improved master equation approach to quantum transport: \\
       From Born to self-consistent Born approximation }

\author{ Jinshuang Jin}
\email{jsjin@hznu.edu.cn}
\affiliation{Department of Physics, Hangzhou Normal University,
          Hangzhou 310036, China}

\author{ Jun Li}
\affiliation{ Beijing Computational Science Research Center,
              Beijing 100084, China }
\affiliation{Department of Physics, Hangzhou Normal University,
          Hangzhou 310036, China}

\author{ Yu Liu}
\affiliation{State Key Laboratory for Superlattices and Microstructures,
         Institute of Semiconductors,
         Chinese Academy of Sciences, Beijing 100083, China}

\author{ Xin-Qi Li}
\email{lixinqi@bnu.edu.cn}
\affiliation{State Key Laboratory for Superlattices and Microstructures,
         Institute of Semiconductors,
         Chinese Academy of Sciences, Beijing 100083, China}
\affiliation{Department of Physics, Beijing Normal University,
Beijing 100875, China}
\affiliation{Department of Chemistry, Hong Kong University
      of Science and Technology, Kowloon, Hong Kong}

\author{YiJing Yan}
\email{yyan@ust.hk}
\affiliation{Department of Chemistry, Hong Kong University
      of Science and Technology, Kowloon, Hong Kong}
\affiliation{Hefei National Laboratory for Physical Sciences at the
 Microscale, University of Science and Technology of China, Hefei,
 Anhui 230026, China}

\date{\today}

\begin{abstract}
Beyond the second-order Born approximation,
we propose an improved master equation approach
to quantum transport under self-consistent Born approximation.
The basic idea is to replace the {\it free} Green's function
in the tunneling self-energy diagram by an {\it effective}
reduced propagator under the Born approximation.
This simple modification has remarkable consequences.
It not only recovers the exact results for quantum transport 
through noninteracting systems under arbitrary voltages, 
but also predicts the challenging nonequilibrium Kondo effect.
Compared to the nonequilibrium Green's function technique that 
formulates the calculation of specific correlation functions,
the master equation approach contains richer dynamical information 
to allow more efficient studies for such as the shot noise 
and full counting statistics.
\end{abstract}



\maketitle

\section{Introduction}

The Landauer-B\"uttiker scattering theory and the nonequilibrium 
Green's function (nGF) approach are widely applied as two 
standard methods for mesoscopic quantum transports \cite{Dat95,Hau96}.
As alternative choices, the classical rate equation \cite{Gla88,Dav93,Naz93} 
and quantum master equation \cite{Gur96a,Gur96b,Sch94,Sch96,Sch06,Li05a,Li05b}
are more convenient in some cases.
In particular, the number-resolved version of the quantum master equation 
approach \cite{Li05a,Li05b,Shn98+01} has been demonstrated very
useful for the study of quantum noise, counting statistics,
and large-derivation analysis \cite{Li11}.

In most cases, such as in quantum optics, the second-order 
master equation (ME) is widely applied and works perfectly. 
However, for quantum transports, the second-order expansion of the tunneling 
Hamiltonian only corresponds to sequential-tunneling-governed transport,
which does not incorporate the level broadening effect,
implying thus a validity condition of {\it large bias voltage}.
Moreover, for interacting systems, despite the second-order ME 
can predict such as the Coulomb staircase behavior,
it cannot deal with the cotunneling and Kondo effects.
To break through this limitation, higher-order expansions 
for the tunneling Hamiltonian are required
\cite{Sch94,Sch96,Sch06,Yan080911,Yan2012,Wac05+10,CS11,
Leeu09,Galp09,Galp10,KG13}.

The second-order master equation is obtained from 
the well-know Born approximation through 
perturbative expansion of the tunneling Hamiltonian.
The resultant {\it dissipation} term, in analogy to
the quantum dissipative system,
corresponds to a {\it self-energy} process of tunneling.
On the other hand, it is well known that in the Green's function theory,
an efficient scheme of higher-order correction
is the use of renormalized self-energy diagram
under the so-called {\it self-consistent Born} approximation (SCBA),
which is actually a type of self-consistent renormalization
to the bare propagator with a {\it dressed} one \cite{Mat}.
From this insight, for quantum transport
we may replace the free (system-Hamiltonian only) Green's function
in the second-order self-energy diagram, with an {\it effective}
propagator defined by the second-order ME.
We will see that the effect of this improvement is remarkable:
it recovers not only the {\it exact} result
of noninteracting transport under {\it arbitrary} voltages,
but also the {\it cotunneling} and nonequilibrium {\it Kondo} 
features for interacting systems.

Similar ideas of modifying the free propagator
in the tunneling self-energy diagram by a dressed one
were implemented also in a couple of recent studies \cite{Galp09,Galp10,KG13}. 
But the master equation of basis-free superoperator form was not obtained,
and Ref.\ \cite{KG13} aimed to a study on the Anderson impurity model
heavily based on a diagrammatic technique.
Moreover, owing to inappropriately treating the dressed propagator
as a Markovian-Redfield generator \cite{Galp09,Galp10}, 
unsatisfactory problems occurred,  
as stated in the concluding remarks of Ref.\ \cite{Galp10}:
`` $\cdots$ Note however that many important effects due to strong correlation 
between the molecule and contacts observed at low temperatures 
(e.g., Kondo) cannot be reproduced within our scheme. 
We find that our scheme becomes unreliable in the region of the parameters 
where coherences in the system eigenbasis 
(i.e., coherences introduced through nondiagonal elements of 
molecule-contact coupling matrix $\Gamma$) are bigger than 
the interlevel separation and on the order of the diagonal elements 
of the molecule-contact coupling matrix $\Gamma$".
Remarkably, our treatment in the present study clears out 
all these unsatisfactory problems.

The paper is organized as follows.
In Sec.\ II we present the main formulation
of the master equation approach under SCBA.
This central part constitutes a number of subsections:
we first outline in Sec.\ II.A the master equation approach to quantum transport
under the Born approximation, then discuss in Sec.\ II.B the basic idea of the SCBA
which is further implemented in Sec.\ II.C to construct
the improved master equation;
subsequently, in Sec.\ II.D and E we consider the steady state
and prove an {\it exact equivalence}
to the nGF approach for noninteracting systems.
In Sec.\ III, we perform a more challenging test
by applying the SCBA-ME to transport through an interacting quantum dot,
where the recovery of the nonequilibrium Kondo effect will be demonstrated.
Finally, we summarize the work in Sec.\ V.
As complementary materials, we arrange five Appendices
for some technical details in addition to the main text.


\section{Self-consistent Born approximation: Formulations}

\subsection{Scheme under Born approximation}

Let us start with a transport setup described by
\be\label{H-ms}
 H = H_S(a_{\mu}^{\dg},a_{\mu})+ H_B + H_{SB} .
\ee
In this Hamiltonian, $H_S$ is for the central {\it system}
embedded between two leads that are regarded as a generalized
{\it environment} and are modeled by
$H_B=\sum_{\alpha={\text{L,R}}}\sum_{k}(\epsilon_{\alpha  k}+\mu_{\alpha})
b^{\dg}_{\alpha\mu k}b_{\alpha k}$, under the bias voltage of
$V=(\mu_{\rm L}-\mu_{\rm R})/e$.
The coupling between the system and the leads
is described by a tunneling Hamiltonian,
$H_{SB}=\sum_{\alpha\mu k}(t_{\alpha\mu k}
a^{\dg}_{\mu}b_{\alpha k}+\rm{H.c.})$.
Here, $a^{\dg}_{\mu}$ and  $b^{\dg}_{\alpha k}$ ($a_{\mu}$ and  $b_{\alpha k}$)
are the electron creation (annihilation) operators of the specified
system and $\alpha$-lead states,
while $\epsilon_{\alpha k}$ and $t_{\alpha\mu k}$ denote the state energy
and the coupling integral parameters, respectively.
Following the standard treatment of quantum open systems,
we introduce a collective reservoir operator,
$F_{\alpha\mu} = \sum_{k} t_{\alpha\mu k}b_{\alpha k}$,
for the coupling between the $\alpha=\text{L,\ R}$ lead
and the system state ``$\mu$''.
We can therefore rewrite the tunneling Hamiltonian as
$H_{SB}= \sum_{\alpha\mu} \left( a^{\dg}_{\mu} F_{\alpha\mu}
       + \rm{H.c.}\right)$.
By expanding $H_{SB}$ perturbatively up to the second-order,
i.e., under the Born approximation,
we obtain a formal master equation expression with memory as \cite{Yan98}:
\be\label{cumm-expan}
 \dot\rho(t)=-i{\cal L} \rho(t)-\int_{t_0}^{t} d\tau
  \Sigma^{(2)}(t-\tau) \rho(\tau).
\ee
Here, the reduced density matrix of the {\it system}, $\rho(t)$,
is defined by tracing out the reservoir states from the entire
system-plus-reservoir density matrix $\rho_T(t)$,
i.e., $\rho(t)=\rm{Tr}_B[\rho_T(t)]$.
Two superoperators,  the system Liouvillian ${\cal L}$,
defined via ${\cal L}(\cdots)= [H_S,(\cdots)]$,
and the second-order self-energy superoperator $\Sigma^{(2)}
= \la {\cal L}'(t) {\cal G}(t,\tau){\cal L}'(\tau)\ra$,
are introduced in \Eq{cumm-expan}.
Defined here are also
the Liouvillian ${\cal L'}(\cdots)=[H_{SB},(\cdots)]$ and the
free (system) propagator ${\cal G}(t,\tau)=e^{-i{\cal L}(t-\tau)}$.
Reexpressing the Liouvillian self-energy superoperator in Hilbert-space,
the integrand in \Eq{cumm-expan} has four terms,
see Appendix A and Fig.\ 1 where a diagrammatic illustration
on the real-time Keldysh contour is employed.
We obtain the master equation (\ref{cumm-expan})
a compact operator form of
\be\label{rhot-second}
  \dot\rho(t) =-i{\cal L}\rho(t)
  - \sum_{\mu\sigma}\Big\{\big[a^{\bar\sigma}_\mu,
  A^{(\sigma)}_{\mu\rho}(t)\big]
  +{\rm H.c.}
   \Big\} .
\ee
Here, for the sake of brevity, we make the following conventions:
$\sigma=+$ and $-$, $\bar\sigma=-\sigma$;
$a^+_\mu = a^\dg_\mu$ and $a^-_\mu = a_\mu$. Introduced in \Eq{rhot-second} is also
$A^{(\sigma)}_{\mu\rho }(t)\equiv\sum_{\alpha} A^{(\sigma)}_{\alpha\mu\rho}(t)$,
where
\be\label{Arho-second}
A^{(\sigma)}_{\alpha\mu\rho}(t)=\sum_\nu\int^t_{t_0} d\tau
C^{(\sigma)}_{\alpha\mu\nu}(t-\tau)
\left\{{\cal G}(t,\tau)[a^{\sigma}_\nu\rho(\tau)]\right\},
\ee
with $ C^{(\sigma)}_{\alpha\mu\nu}(t-\tau)
= \la F^{(\sigma)}_{\alpha\mu}(t)F^{(\bar\sigma)}_{\alpha \nu}(\tau) \ra_{\rm B}$.
The time dependence in $F^{(\pm)}_{\alpha\mu}(t)$
originates from the interaction picture
with respect to the Hamiltonian of the leads,
and the average $\la \cdots \ra_B$ is over the lead states.
It should be noted that, in deriving the above results,
only the Born approximation was used,
but not involving the Markovian approximation.
The non-Markovian feature is reflected by the time {\it non-local}
self-energy terms in \Eq{Arho-second}.

Finally, following Ref.\ \cite{Li05b},
the transport current is given by
\be\label{It-second}
I_{\alpha}(t)=\frac{2e}{\hbar}\sum_\mu {\rm Re}
\left\{ {\rm Tr}\big[ A^{(+)}_{\alpha\mu\rho}(t)a_\mu -
 A^{(-)}_{\alpha\mu\rho}(t) a^\dg_\mu\big] \right\}.
\ee
Here the trace is over the states of the central system.

\subsection{Basic consideration}
\label{thsec2B}

Strictly speaking, the second-order master equation applies only to transport
under large bias voltage. That is, the Fermi levels of the leads
should be considerably far away from the transport levels of the central system,
being at least several times of the level's broadening.
This can be understood by the simplest example
of resonant transport through a single-level quantum dot.
For this simple system, one can prove that the second-order master equation
will result in a vanishing current at zero temperature,
if the dot level $E_0$ is located slightly outside the bias window.
However, it is well known that a full quantum mechanical treatment
will give a nonzero tunneling current in this situation.
Similar difficulty arises as well if the dot level $E_0$
is in between the Fermi levels, ($\mu_{\rm L}>E_0>\mu_{\rm R}$),
the second-order theory will always predict a full resonant current
of $I=e\Gamma_{\rm L}\Gamma_{\rm R}/(\Gamma_{\rm L}+\Gamma_{\rm R})$,
no matter how small the Fermi levels are away from $E_0$.
Insightfully, these unreasonable results are
associated with the neglect of the level's broadening effect.

We notice that the tunneling self-energy operator
in \Eq{cumm-expan},
$\Sigma^{(2)}(t-\tau) = \la {\cal L}'(t) {\cal G}(t,\tau){\cal L}'(\tau)\ra$,
contains a free (system only) Green's function ${\cal G}(t,\tau)$,
which in the case of single-level dot reads ${\cal G}(t,\tau)\sim e^{-iE_0(t-\tau)}$.
Then, if we could attach $e^{-\Gamma |t-\tau|}$ to this unitary
propagator, the level broadening effect would be restored.
Physically, this corresponds to a certain {\it self-consistent Born}
correction to the tunneling self-energy.
Moreover, as to be shown in the following, the improvement
from {\it Born} to {\it self-consistent Born} approximation
to the tunneling self-energy contains more than the broadening effect.
In general, it includes also an energy shift and, moreover,
the interplay between the coherent multiple tunneling and the on-site strong
Coulomb interaction which are essential to the Kondo effect.

\begin{figure}
\includegraphics[width=8cm]{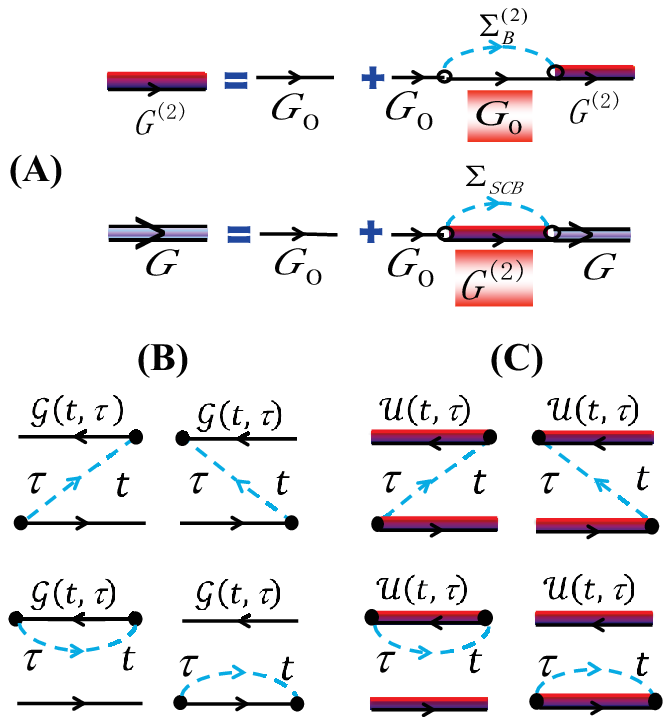}
\caption{(color online) (A): The self-consistent Born approximation
in Green's function theory, where the free Green's function $G_0$ in
the self-energy is replaced by an effective one, $G^{(2)}$. The
dashed curve represents, for instance, the Green's function of
phonon/photon in the case of an electron-phonon/photon interacting
system. 
(B): The four second-order tunneling self-energy diagrams,
$\Sigma^{(2)}(t-\tau)$, in the real-time Keldysh contour. The dashed
lines are the Green's functions of the reservoir electrons. 
(C): The improved tunneling self-energy diagrams under the
self-consistent Born approximation, in which the free (system only)
Green's function ${\cal G}(t,\tau)$ was replaced by the second-order
effective propagator ${\cal U}(t,\tau)$.    }
\end{figure}

 In the Green's function theory, it is well known that
the correction of the self-energy diagram
under self-consistent Born approximation (SCBA)
is an efficient scheme for a partial inclusion of
high-order tunneling contributions.
In Fig.\ 1(A), taking the electron-phonon (or photon) interaction
as an example, we illustrate the main consideration here.
The solid line is for the free Green's function $G_0$
of the electron, the double lines for the full Green's function $G$,
and the dashed line for the Green's function of the phonon (or photon).
The basic idea of the self-consistent Born approximation is replacing
$G_0$ in the self-energy diagram by an improved one, $G^{(2)}$,
as shown in Fig.\ 1(A).
This correction corresponds to an infinite
re-summation of the second-order Born self energy
and, in most cases, can largely improve the results.

  The Keldysh diagrammatic representation of
the tunneling self-energy
$\Sigma^{(2)}= \la {\cal L}'(t) {\cal G}(t,\tau){\cal L}'(\tau)\ra$
is shown in Fig.\ 1(B).
It involves the unperturbed bare system propagator, ${\cal G}(t,\tau)$.
Thus, any vertical line between vertexes in Fig.\ 1(B) crosses one tunneling
line (the dashed line), leading to the lowest-order perturbation
master equation described by \Eq{cumm-expan} or
(\ref{rhot-second}).
In the spirit of SCBA the modified version
assumes $\Sigma_\text{\tiny SCB}
= \la {\cal L}'(t) {\cal U}(t,\tau){\cal L}'(\tau)\ra$,
with the propagator ${\cal U}(t,\tau)$ arising from
the ME (\ref{cumm-expan}) that incorporates $\Sigma^{(2)}$.
Consequently, a vertical line between vertexes in
$\Sigma_\text{\tiny SCB}$ of Fig.\ 1(C) involves {\it effectively},
on top of the single-tunneling crossing in Fig.\ 1(B), also the
double-tunneling crossing diagram.
The resultant SCBA-ME propagator acquires the desired
iteration nature, incorporating therefore
infinite tunneling processes \cite{Sch94,Sch96}.
The underlying self-consistent
$\Sigma_\text{\tiny SCB}$ re-summation scheme surely
includes multiple interaction lines
connecting the horizontal propagations
without intersection \cite{Sch94,Sch96}.
As the self-energy is treated at the master equation level,
the basic requirement of probability conservation,
i.e., $\frac{d}{dt}{\rm Tr}\rho(t)=0$, is always
preserved [cf.\ \Eq{QME-SCBA}].

 A crucial issue in developing a SCBA-ME
for electronic transport systems is the
Fermi-Grassmann parity difference between
$\Sigma^{(2)}$ and $\Sigma_\text{\tiny SCB}$ (cf.\ \Sec{thsec2C}).
This issue appears also in the hierarchical equation of motion (HEOM)
formalism \cite{Yan080911} and the real-time diagrammatic (RTD) technique  \cite{Sch96}.
For further comparison, we may symbolically represent
the self-energy in frequency domain as
$\Sigma_\text{\tiny SCB}(\omega)\sim
{\cal L}'[\omega - {\cal L}+i\Sigma^{(2)}(\omega)]^{-1}{\cal L}'$.
It highlights the local-frequency dependence of $\Sigma_\text{\tiny SCB}(\omega)$
on $\Sigma^{(2)}(\omega)$.
In other words, $\Sigma_\text{\tiny SCB}$ of Fig.\ 1(C)
does not access the horizontal intersections of nonlocal frequencies.
As inferred from the second-tier-level HEOM formalism (cf.\ \App{app_HEOM})
that is equivalent to the RTD approach \cite{Yan080911,Sch96},
each individual horizontal intersection
could be of the same order
as those vertical intersections treated in Fig.\ 1(C).
In this regard the present SCBA scheme resembles
a random-phase approximation, which assumes
the integrated contribution from all nonlocal frequencies
be negligible, in comparing to those local-frequency contributions.
Remarkably, the approximation here is truly valid,
as supported by the resulting steady-state properties
the following two observations:
(\emph{i}) For noninteracting transport systems
the present SCBA scheme is \emph{exact} and
recovers the second-tier-level HEOM results
 (cf.\ \Sec{thsec2D});
(\emph{ii}) For interacting systems it reproduces nGF equation-of-motion (EOM)
results \cite{Hau96} including those of nonequilibrium
Kondo features [cf.\ \Eq{Kondo}].
The above observations conclude
that the SCBA-ME in this work,
despite of its random-phase simplification,
does treat the vertex and the self energy at
the same level of approximation.
It renders an \emph{efficient} and \emph{reliable}
means for various quantum transport problems,
including nonequilibrium Kondo cotunnelings in
interacting systems.
In the coming subsections, we present the SCBA-ME,
with the details on the aforementioned
features and observations.

\subsection{Scheme under self-consistent Born approximation}
\label{thsec2C}

Based on \Eq{rhot-second}
we formally introduce an evolution operator ${\cal U}(t,\tau)$,
which propagates the state in terms of
$\rho(t)={\cal U}(t,\tau)\rho(\tau)$.
 Then, based on the insight above, we replace
${\cal G}(t,\tau)$ with ${\cal U}(t,\tau)$
in the self-energy operator or more precisely
in $A^{(\sigma)}_{\alpha\mu\rho}$ [c.f. \Eq{Arho-second}]:
\begin{align}\label{Arho-SCBA}
{\cal A}^{(\sigma)}_{\alpha\mu\rho}(t)
&=\sum_\nu\int^t_{t_0} d\tau C^{(\sigma)}_{\alpha\mu\nu}(t-\tau)
\left\{{\cal U}(t,\tau)[a^{\sigma}_\nu\rho(\tau)]\right\} .
\end{align}
Inserting this improved quantity into the master equation
and the transport current, we have
\be \label{QME-SCBA}
  \dot\rho(t) =-i{\cal L}\rho(t)
  - \sum_{\mu\sigma}\Big\{\big[a^{\bar\sigma}_\mu,
  {\cal A}^{(\sigma)}_{\mu\rho}(t)\big]
  +{\rm H.c.}
   \Big\} ,
\ee
where
${\cal A}^{(\sigma)}_{\mu\rho}(t)\equiv \sum_{\alpha}{\cal A}^{(\sigma)}_{\alpha\mu\rho}(t)$,
and
\be\label{It-SCBA}
   I_{\alpha}(t)=\frac{2e}{\hbar}\sum_\mu {\rm Re}
\left\{ {\rm Tr}\big[ {\cal A}^{(+)}_{\alpha\mu\rho}(t)a_\mu
 -{\cal A}^{(-)}_{\alpha\mu\rho}(t)a^\dg_\mu\big] \right\}.
\ee
Desirably, \Eqs{QME-SCBA} and (\ref{It-SCBA}) have the same
compact structures as \Eqs{rhot-second} and (\ref{It-second}),
respectively, in the second-order Born master equation approach.
The only difference is the replacement of $A^{(\pm)}_{\alpha\mu\rho}(t)$
of \Eq{Arho-second}
by ${\cal A}^{(\pm)}_{\alpha\mu\rho}(t)$ of \Eq{Arho-SCBA}.
The most obvious consequence of this improvement is that
the {\it broadening effect} and {\it energy shift}
induced by the tunneling are included
in the system state evolution in the self-energy terms.
But, not only limited to these, it has even more implications.
For instance, a careful inspection of Fig.\ 1(C) reveals that
this replacement, significantly, accounts for the interplay
of the multiple tunneling processes
and the Coulomb interactions inside the central system.
It is well known that such type of interplay is the key reason
for Kondo effect, including the nonequilibrium Kondo effect
in transport though the Anderson-type impurities.
In Sec.\ III we will detail an example for this issue.
Moreover, the cotunneling processes are also
most naturally contained in the proposed SCBA-ME scheme.
Under the bias condition of Coulomb blockade, the SCBA-ME can recover
the cotunneling results given by other approaches \cite{Sch06}.  

Below we outline a protocol to solve \Eq{QME-SCBA} in frequency domain,
by the Laplace transformation
$\rho(\omega)=L[\rho(t)] = \int^{\infty}_{0}dt \,e^{i\omega t} \rho(t)$.
Accordingly, \Eq{QME-SCBA} reads
\begin{align} \label{rhow-SCBA}
& -i\omega\rho(\omega) -\rho(0)
 = -i {\cal L}\rho(\omega)  \nl
& - \sum_{\mu\sigma}
 \left\{ \big[a^{\bar\sigma}_{\mu},{\cal A}_{\mu\rho}^{(\sigma)}(\omega)\big]
 - \big[a^{\sigma}_{\mu},{\cal A}_{\mu\rho}^{(\sigma)\dg}(-\omega)\big] \right\} ,
\end{align}
where ${\cal A}_{\mu\rho}^{(\sigma)}(\omega)
=\sum_{\alpha} {\cal A}_{\alpha\mu\rho}^{(\sigma)}(\omega)$,
with ${\cal A}_{\alpha\mu\rho}^{(\sigma)}(\omega)$
explicitly expressed as
\begin{align}\label{Arhow-SCBA}
{\cal A}^{(\pm)}_{\alpha\mu\rho}(\omega)
& =
\sum_\nu\int^\infty_{-\infty} \frac{d\omega'}{2\pi}
\Gamma^{(\pm)}_{\alpha\mu\nu}(\omega')
{\cal U}(\omega\pm\omega')[a^{\pm}_\nu\rho(\omega)].
\end{align}
In deriving this result that is limited to the case of
${\cal U}(t,\tau)={\cal U}(t-\tau)$, we
have used the simple relation,
$L[e^{\pm i\omega' t} {\cal U}(t)]={\cal U}(\omega\pm\omega')$,
and introduced the Fourier expansion
$ C^{(\pm)}_{\alpha\mu\nu}(t)=\int \frac{d\omega}{2\pi}
e^{\pm i\omega t}\Gamma^{(\pm)}_{\alpha\mu\nu}(\omega)$.
More explicitly,
$\Gamma^{(+)}_{\alpha\mu\nu}(\omega)
=\Gamma_{\alpha\nu\mu}(\omega)f^{(+)}_\alpha(\omega)$,
and $\Gamma^{(-)}_{\alpha\mu\nu}(\omega)
=\Gamma_{\alpha\mu\nu}(\omega)f^{(-)}_\alpha(\omega)$,
where $\Gamma_{\alpha\mu\nu}(\omega)=2\pi\sum_{k}t_{\alpha\mu k}t^\ast_{\alpha\nu k}
\delta(\omega-\epsilon_{\alpha k})$ is the spectral density function
of the $\alpha$-lead.
The other two quantities, $f^{(+)}_\alpha(\omega)=f_\alpha(\omega)$
and $f^{(-)}_\alpha(\omega)=1-f_\alpha(\omega)$,
are the occupied and unoccupied Fermi functions, respectively.
Equations (\ref{rhow-SCBA}) and (\ref{Arhow-SCBA}) constitute
 a closed form of master equation in frequency domain,
which allows for a straightforward way to get the solution.
However, in doing this, we must explicitly identify the propagator
resolution
${\cal U}(\omega)$ in \Eq{Arhow-SCBA}.
More specifically, we need to consider the evolution of
$\ti\rho_j(t)\equiv {\cal U}(t,\tau)[a^{\sigma}_\nu\rho(\tau)]$,
where we use a single index ``$j$'' to denote
``$\{{\nu,\sigma}\}$'' for brevity.

 Care must be taken in treating the reduced propagator
${\cal U}(t,\tau)$ in \Eq{Arho-SCBA},
as it involves the issue of Grassman-Fermi's parity.
Originally, the second-order reduced propagator ${\cal U}$
was introduced via the usual propagation
of a {\it physical} state (density matrix),
i.e., $\rho(t)={\cal U}(t,t_0)\rho(t_0)$.
However, in \Eq{Arho-SCBA} or (\ref{Arhow-SCBA}),
the quantity being propagated is $a^{\sigma}_\nu\rho$,
which has the different Grassmannian parity from
the density matrix.
Conventionally, one may expect that
the propagator would be {\it independent} of the quantity to be propagated.
However, the analysis in Appendix A shows that this ``general'' rule
breaks down quite unexpectedly in the present case.
The basic reason is that the quantity to be propagated,
$a^{\sigma}_\nu\rho$, contains an extra fermionic electron
operator, in compared to the density operator itself.
Owing to the Pauli principle,
an extra minus sign would appear in two of the four
self-energy terms in its equation of motion.
This converts the {\it commutators} in the usual master equation
to the {\it anti-commutators}; see \Eq{tirhoj} below.
The involving details are reported in Appendix A.    
We will find that, through the illustrative examples of this work,
this subtle issue is crucially important for
the present theory to have the correct propertied presented later in this work.

We summarize the result derived in \App{thapp-propa}
for $\ti\rho_j(t)\equiv {\cal U}(t,\tau)[a^{\sigma}_\nu\rho(\tau)]$ as follows:
\begin{align}\label{tirhoj}
  \dot{\ti\rho}_j(t) &=-i{\cal L}\ti\rho_j(t)
 -\sum_{\mu}
\Big[\big\{a_\mu,A^{(+ )}_{\mu\ti\rho_j}\big\}
+\big\{a^\dg_\mu,A^{(-)}_{\mu\ti\rho_j}\big\}
\nl&\quad+
\big\{a^\dg_\mu,A^{(+ )\dg}_{\mu\ti\rho_j}\big\}
+\big\{a_\mu,A^{(- )\dg}_{\mu\ti\rho_j}  \big\} \Big].
\end{align}
The operators $A^{(\pm)}_{\mu\ti\rho_j}$ in this equation
have the same form of $A^{(\pm)}_{\mu\ti\rho}$
in \Eq{Arho-second}, needing only to replace $\rho$ by $\ti\rho_j$.
As emphasized earlier, a significant difference
appears \emph{unexpectedly} between \Eq{tirhoj} and \Eq{rhot-second}.
That is, the {\it commutators} in the master equation
(\ref{rhot-second}) become now the {\it anti-commutators} in \Eq{tirhoj}. 

In frequency domain, the solution of \Eq{tirhoj}
determines the propagator resolution ${\cal U}(\omega)$ in \Eq{Arhow-SCBA}.
The Laplace transform of \Eq{tirhoj},
\begin{align} \label{tirhow}
-i\omega\ti\rho_j(\omega)-\ti\rho_j(0)
& = -i {\cal L}\ti\rho_j(\omega) - \Sigma(\omega)\ti\rho_j(\omega),
\end{align}
gives
\be\label{Uomega}
{\cal U}(\omega) = \left[ i({\cal L}-\omega)+\Sigma(\omega) \right]^{-1}.
\ee
The involving self-energy superoperator in frequency domain reads
\begin{align}\label{Sigmaw}
\Sigma(\omega)&=
 \!\sum_{\sigma\mu\nu}\!
   \Big[\leftact{a}\,\!^{\bar\sigma}_{\mu}
     C^{(\sigma)}_{\mu\nu}(\omega\!-\!{\cal L})
   \leftact{a}\,\!^{\sigma}_{\nu}
  +\!
   \rightact{a}\,\!^{\sigma}_{\mu}
     C^{(\sigma)\ast}_{\mu\nu}({\cal L}\!-\!\omega)
   \rightact{a}\,\!^{\bar\sigma}_{\nu}
\nl&\quad\
   +\! \rightact{a}\,\!^{\bar\sigma}_{\mu}
    C^{(\sigma)}_{\mu\nu}\!(\omega\!-\!{\cal L})
   \leftact{a}\,\!^{\sigma}_{\nu}
  +\!
   \leftact{a}\,\!^{\sigma}_{\mu}
       C^{(\sigma)\ast}_{\mu\nu}\!({\cal L}\!-\!\omega)
   \rightact{a}\,\!^{\bar\sigma}_{\nu} \Big].
\end{align}
The shorthand notations introduced here,
$\leftact{a}\,\!^{\sigma}_{\mu}\hat O \equiv {a^{\sigma}_{\mu}}\hat O$
and
$\rightact{a}\,\!^{\sigma}_{\mu}\hat O \equiv \hat O {a^{\sigma}_{\mu}}$,
considerably simplify the expression.
And, $C^{(\sigma)}_{\alpha\mu \nu}(\omega)$,
the Laplace transformation of $C^{(\sigma)}_{\alpha\mu \nu}(t)$,
are related with $\Gamma^{(\pm)}_{\alpha\mu \nu}(\omega)$
through the well known dispersive relation:
\begin{align}\label{FDT2}
C^{(\pm)}_{\alpha\mu \nu}(\omega)
&=\int^\infty_{-\infty}\frac{d\omega'}{2\pi}
\frac{i}{\omega\pm\omega'+i0^+}\Gamma^{(\pm)}_{\alpha\mu \nu}(\omega').
\end{align}

 Equations (\ref{Arho-SCBA}), (\ref{QME-SCBA}) and (\ref{tirhoj})
constitute the central formulation of the SCBA-ME.
In Appendix D we prove a relation of the SCBA-ME
with an alternative approach, say, the (infinite)
hierarchical equations of motion (HEOM) approach \cite{Yan080911}.
The latter is derived by a series of derivatives on the
Feynman-Vernon influence functional,
based on a spectral decomposition technique
and introducing a series of auxiliary operators.
In Appendix D we show that the SCBA-ME
contains the dominant contribution of the HEOM formulation.  

\subsection{ Steady state }
\label{thstationary}

 Based on \Eqs{rhow-SCBA}, (\ref{Arhow-SCBA}) and (\ref{tirhow})
one can first carry out the solution in frequency domain.
Then, by an inverse Laplace transformation,
the time-dependent solution can be obtained.
Nevertheless, in this subsection we would like to show
an efficient scheme for the steady state solution.  
For this purpose, let us consider the integral in \Eq{Arho-SCBA}.
Since the correlation function in the integrand
of \Eq{Arho-SCBA} is nonzero only on {\it finite} timescale,
we can replace $\rho(\tau)$
in the integrand by the steady-state density matrix $\bar{\rho}$,
in the long time limit $t\rightarrow\infty$
(corresponding to the steady state).
After this, we first make a Fourier expansion for
$C^{(\sigma)}_{\alpha\mu\nu}(t-\tau)$, then perform the Laplace transform
for ${\cal U}(t-\tau)[a^{\pm}_\nu \bar{\rho}]$, yielding
\begin{align}\label{SupA_st}
{\cal A}^{(\pm)}_{\alpha\mu\bar\rho}(t\rightarrow\infty)
&=\sum_\nu\int^\infty_{-\infty}\frac{d\omega}{2\pi}\,
\Gamma^{(\pm)}_{\alpha\mu\nu}(\omega)
{\cal U}(\pm\omega)[a^{\pm}_\nu\bar\rho] .
\end{align}
Together with \Eq{tirhow}, substituting this result into \Eq{QME-SCBA},
we can straightforwardly solve for $\bar\rho$
and calculate the steady-state current.

To get further insight into the SCBA-ME scheme,
we recast the current formula \Eq{It-SCBA}
in a more conventional form.
To this end we introduce:
$\varphi_{1\mu\nu}(\omega)={\rm Tr} \big[a_\mu\ti\rho_{1\nu}(\omega)\big]$,
and
$\varphi_{2\mu\nu}(\omega)={\rm Tr} \big[a_\mu\ti\rho_{2\nu}(\omega)\big]$,
where $\ti\rho_{1\nu}(\omega)$ and $\ti\rho_{2\nu}(\omega)$
are the solution of \Eq{tirhow}, with an initial condition of
$\ti\rho_{1\nu}(0)=\bar\rho a^\dg_\nu$ and
$\ti\rho_{2\nu}(0)=a^\dg_\nu\bar\rho $.
Moreover, we introduce a matrix notation using, for instance,
$\bm \varphi_j$ to denote $\varphi_{j\mu\nu}$.
Then, the steady-state current via \Eq{It-SCBA}
can be expressed in a compact form as
\begin{align}\label{Is-SCBA}
   \bar I_{\alpha}=
   \frac{2e}{\hbar}{\rm Re}\!
   \int^\infty_{-\infty}\!\!\frac{d\omega}{2\pi}{\rm Tr}\big\{
   \bm\Gamma_{\alpha}(\omega)\left[f_\alpha(\omega) \bm\varphi(\omega)
    -  \bm\varphi_1(\omega)\right]\!\big\} ,
\end{align}
where $\bm\varphi(\omega)=\bm\varphi_1(\omega)+\bm\varphi_2(\omega)$.

Further simplification is possible, if $\bm\Gamma_{\rm L}=\lambda\bm\Gamma_R$,
where $\lambda$ is a constant.
In this case, \Eq{Is-SCBA} can be recast to the Landauer-B\"uttiker type
of current formula. That is, the current is an integration of
tunneling coefficient over the bias window:
$ \bar I = \frac{2e}{\hbar}{\rm Re}
   \int^\infty_{-\infty}\frac{d\omega}{2\pi}
    \left[ f_{\rm L}(\omega)- f_{\rm R}(\omega)\right] {\cal T}(\omega) $.
In our case, the effective tunneling coefficient reads
$  {\cal T}(\omega)={\rm Tr}\{
  \bm\Gamma_{\rm L}\bm\Gamma_{\rm R}
  (\bm\Gamma_{\rm L}+\bm\Gamma_{\rm R})^{-1}
  {\rm Re}\big[\bm\varphi(\omega)\big]\}. $
Compared to the nGF formulation \cite{Hau96}, we find that
$\bm\varphi$ plays a role of the retarded Green's function,
i.e., $\bm\varphi(\omega)=i\bm G^r(\omega)$.
The point is that, the current formula in terms of the nGF
is only a formal expression: it does not say anything
about the methods to obtain the various Green's functions.
Our $\bm\varphi$, however, is based on a concrete
computational scheme in terms of master equation.
In the following, we shall demonstrate: {\it (i)} the SCBA-ME approach
is {\it exact} for noninteracting transport under arbitrary voltage;
and {\it (ii)} it is likely to be good enough for interacting systems
-- it can predict, for instance, the nonequilibrium Kondo effect.

\subsection{Noninteracting system: Recovery of the exact
            result under arbitrary bias voltage}
\label{thsec2D}

Consider the transport through a noninteracting system:
\be\label{Hs0}
H_S=\sum_{\mu\nu}h_{\mu\nu}a^\dg_\mu a_\nu.
\ee
Straightforwardly, based on \Eq{tirhow},
we obtain the equation of motion for $\bm \varphi_i$ as
follows (see \App{thapp-varhpiw0} for details):
\begin{align}\label{varphiw0}
-i\omega\bm\varphi_i(\omega)-\bm\varphi_i(0)=-i\bm h\bm\varphi_i(\omega)
-i\bm\Sigma_{0}(\omega)\bm\varphi_i(\omega).
\end{align}
Here and in some other parts of this work we use
the bold face operators to denote the matrices in the
eigenstate representation of the central system Hamiltonian.
In \Eq{varphiw0} $\bm\varphi_i(0)$ stands for the initial condition,
$\varphi_{1\mu\nu}(0)={\rm Tr}\big[a_\mu\bar\rho a^\dg_{\nu}\big]$
and $\varphi_{2\mu\nu}(0)={\rm Tr}\big[a_\mu a^\dg_{\nu}\bar\rho\big]$.
The self-energy operator $\bm\Sigma_{0}$ corresponds to
$ \Sigma_{0\mu\nu}(\omega)
= -i\sum_{\alpha} \big[ C^{(-)}_{\alpha\mu\nu}(\omega)
+ C^{(+)\ast}_{\alpha\mu\nu}(-\omega)\big]$, or more explicitly,
\begin{align}\label{Self1}
\Sigma_{0\mu\nu}(\omega)
&=\int^\infty_{-\infty} \frac{d\omega'}{2\pi}
\frac{\Gamma_{\mu\nu}(\omega')}{\omega-\omega' +i0^+} .
\end{align}
Then, based on \Eq{varphiw0}, summing up $\bm\varphi_1(\omega)$
and $\bm\varphi_2(\omega)$ yields
\be\label{phiomega}
\bm\varphi(\omega)
=i\big[ \omega-\bm h-\bm\Sigma_{0}(\omega)\big]^{-1}
=i\bm G^r(\omega).
\ee
In deriving this result, the cyclic property under trace
and the anti-commutative relation,
$\{a_\mu,a^\dg_\nu\}=\delta_{\mu\nu}$, have been used.

Equation (\ref{phiomega}) is the exact Green's function for transport
through a noninteracting system.
We then conclude that the SCBA-ME approach
is {\it exact} for noninteracting transports.
Inserting \Eq{phiomega} into the current formula,
we can evaluate the current for arbitrary bias voltage.
Therefore, quite desirably, this achievement goes beyond
the usual second-order master equation approach, which is
applicable only in large bias limit even for noninteracting systems.

We would like to reemphasize that the results of
Eqs.\ (\ref{varphiw0})-(\ref{phiomega})
cannot be obtained from the second-order Born approximation.
The basic reason is that, under the second-order Born approximation,
the self-energy terms in Eqs.\ (\ref{tirhow}),
(\ref{varphiw0}) and (\ref{phiomega}) are absent.
Then, the current formula of \Eq{Is-SCBA} is reduced to
the {\it integrated} one under {\it large bias},
even in the case of near-resonance small bias voltage.
This is the difficulty of lacking the ``broadening effect'',
resulting in essentially an average (trace)
of two electron operators over the state density matrix
given by the second-order Born master equation \cite{Li05b}.       


\section{Transport through an interacting quantum dot}
\label{thitd}

 Below we perform a more challenging test on the SCBA-ME,
by considering the transport through a strongly interacting quantum dot.
This system can be modeled by the well-known Anderson impurity Hamiltonian:
\be\label{Ands-H}
H_S = \sum_{\mu}\left(\epsilon_{\mu} a_{\mu}^{\dg}a_{\mu}
       +\frac{U}{2}n_{\mu}n_{\bar{\mu}}\right) .
\ee
Here, the index $\mu$ labels the spin up (``$\uparrow$'') and spin
down (``$\downarrow$'') states, and $\bar{\mu}$ corresponds to the
opposite spin orientation.
The spin-dependent energy level,
$\epsilon_{\mu}$, may account for the Zeeman splitting in the presence of
magnetic field ($B$),
$\epsilon_{\uparrow,\downarrow}=\epsilon_0\pm g\mu_B B$.
In this context, $\epsilon_0$ is the degenerate dot level
in the absence of magnetic field,
whereas $g$ and $\mu_B$ are the Lande-$g$ factor and the Bohr's magneton, respectively.
In the interaction part, say, the Hubbard term
$Un_{\uparrow}n_{\downarrow}$,
$n_{\mu}=a^{\dg}_{\mu}a_{\mu}$ is the number operator
and $U$ represents the interacting strength.
Owing to the existence of this term,
we are unable to obtain a closed equation
for $\bm\varphi_i$ as \Eq{varphiw0} for the noninteracting system.
Alternatively, we search for the steady-state solution
of the superoperator ${\cal A}^{(\sigma)}_{\mu\rho}$
which is the key quantity for the current \Eq{It-SCBA}.  

When applying the SCBA-ME approach to this system,
we notice that the correlation function, $C^{(\pm)}_{\alpha\mu\nu}$,
is diagonal with respect to the spin states, i.e.,
$C^{(\pm)}_{\alpha\mu\nu}(t)=\delta_{\mu\nu}C^{(\pm)}_{\alpha\mu}(t)$,
and
$\Gamma^{(\pm)}_{\alpha\mu\nu}=\Gamma^{(\pm)}_{\alpha\mu}\delta_{\mu\nu}$.
Also, we specify the Hilbert space
by the four states involved in the transport:
$|0\ra$, $|\up\ra$, $|\down\ra$ and $|d\ra$, corresponding to
the empty, spin-up, spin-down and double occupancy states, respectively.
Using this basis, we can reexpress the electron operator
by projection operator as
$a^\dg_\mu=|\mu\ra\la 0|+(-1)^{\mu}|d\ra\la \bar \mu|$,
where the convention $(-1)^{\up(\down)}=+(-) 1$ is implied.    

\begin{figure}
\includegraphics[width=8cm]{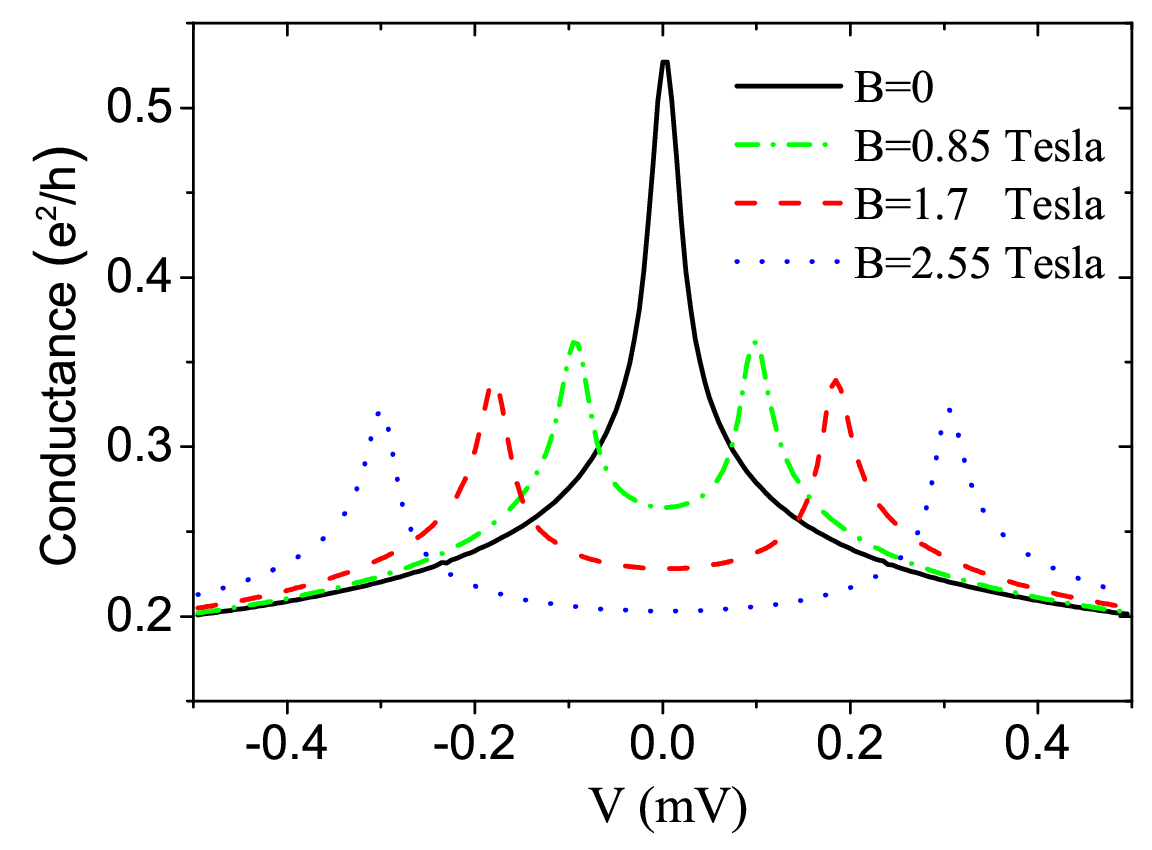}
\caption{(color online)
Kondo peaks in the differential conductance, where the magnetic
field is introduced to generate Zeeman splitting,
$\epsilon_{\uparrow,\downarrow}=\epsilon_0\pm g\mu_B B$.
The adopted parameters follow the experiment in Ref.\ \cite{Ral94}:
the temperature $k_BT =0.005$ meV,
the dot level in the absence of magnetic field $\epsilon_0=-5.2$ meV
(here we take the zero-bias Fermi level as energy reference),
and the on-site Coulomb interaction $U = 120$ meV.
We also consider an Lorentzian spectral density for the (identical) leads
as explained in Appendix C, and assume $\Gamma=3.4$ meV and $W=100$ meV.   }
\end{figure}

For a solution of the steady state, as \Eq{SupA_st}, we have
\begin{align}\label{Arhost-Ad}
{\cal A}^{(\pm)}_{\alpha\mu\bar\rho}
&=\int^\infty_{-\infty}\frac{d\omega}{2\pi}\,
\Gamma^{(\pm)}_{\alpha\mu}(\omega)
{\cal U}(\pm\omega)[a^{\pm}_\mu\bar\rho].
\end{align}
Based on \Eqs{Uomega}-(\ref{FDT2}),
after some algebra (see \App{thapp-III} for more details)
we obtain an analytic expression for
${\cal U}(\pm\omega)[a^{\pm}_\mu\bar\rho]$ as
\be\label{calUw}
\begin{split}
{\cal U}(\omega)[a^{\dg}_\mu\bar\rho]
&=\left[\lambda^+_\mu(\omega)|\mu\ra\la 0|
+\kappa^+_\mu(\omega)(-1)^{\mu}|d\ra\la \bar \mu|\right] ,
\\
{\cal U}(-\omega)[a_\mu\bar\rho]&=\left[\lambda^-_\mu(\omega)|0\ra\la \mu|
+\kappa^-_\mu(\omega)(-1)^{\mu}|\bar \mu\ra\la d|\right] ,
\end{split}
\ee
where
\be\label{lakaw}
\begin{split}
\lambda^+_\mu(\omega)&=i\frac{ \Pi^{-1}_{1\mu}(\omega)\bar\rho_{00}
-\Sigma^-_{\bar\mu}(\omega)\bar\rho_{\bar\mu\bar\mu}}
{ \Pi^{-1}_{\mu}(\omega)\Pi^{-1}_{1\mu}(\omega)
-\Sigma^{+}_{\bar\mu}(\omega)\Sigma^{-}_{\bar\mu}(\omega)},
\\
\lambda^-_\mu(\omega)&=-i\frac{ \Pi^{-1}_{1\mu}(\omega)\bar\rho_{\mu\mu}
+\Sigma^-_{\bar\mu}(\omega)\bar\rho_{dd} }
{ \Pi^{-1}_{\mu}(\omega)\Pi^{-1}_{1\mu}(\omega)
-\Sigma^{+}_{\bar\mu}(\omega)\Sigma^{-}_{\bar\mu}(\omega)},
\\
\kappa^+_\mu(\omega)&=i\frac{ -\Sigma^+_{\bar\mu}(\omega)\bar\rho_{00}
+\Pi^{-1}_{\mu}(\omega)\bar\rho_{\bar\mu\bar\mu} }
{ \Pi^{-1}_{\mu}(\omega)\Pi^{-1}_{1\mu}(\omega)
-\Sigma^{+}_{\bar\mu}(\omega)\Sigma^{-}_{\bar\mu}(\omega)},
\\
\kappa^-_\mu(\omega)&=i\frac{ -\Sigma^+_{\bar\mu}(\omega)\bar\rho_{\mu\mu}
+\Pi^{-1}_{\mu}(\omega)\bar\rho_{dd} }
{ \Pi^{-1}_{\mu}(\omega)\Pi^{-1}_{1\mu}(\omega)
  -\Sigma^{+}_{\bar\mu}(\omega)\Sigma^{-}_{\bar\mu}(\omega)},
\end{split}
\ee
with $\Pi^{-1}_{ \mu}(\omega)=\omega-\epsilon_\mu-\Sigma_{0\mu}(\omega)
 -\Sigma^+_{\bar\mu}(\omega)$ and
$\Pi^{-1}_{1\mu}(\omega)=\omega-\epsilon_\mu-U-\Sigma_{0\mu}(\omega)
-\Sigma^-_{\bar\mu}(\omega)$.
The self-energy $\Sigma_{0\mu}(\omega)$ is given by \Eq{Self1},
while $\Sigma^{\pm}_{\mu}(\omega)$ is defined through
 \begin{align}
\Sigma^{\pm}_{\mu}(\omega)&=\int^{\infty}_{-\infty}\! \frac{d\omega'}{2\pi}
\frac{\Gamma^{(\pm)}_{\mu}(\omega')}{\omega-\epsilon_{\bar\mu}
+\epsilon_\mu-\omega' +i0^+}
\nl&\quad
+\int^{\infty}_{-\infty}\! \frac{d\omega'}{2\pi}
\frac{\Gamma^{(\pm)}_{\mu}(\omega')}{\omega-E_d+\omega' +i0^+}.
\end{align}
Denote further
$\Sigma_{\mu}(\omega)\equiv \Sigma^{+}_{\mu}(\omega)+\Sigma^{-}_{\mu}(\omega)$.
Then, we find the solution of $\varphi_{\mu\mu}(\omega)$ as
\begin{align}\label{Kondo}
&\ \varphi_{\mu\mu}(\omega)
=\frac{i\big[\Pi^{-1}_{1\mu}(\omega)
+\Sigma^{+}_{\bar\mu}(\omega)\big]
(1-n_{\bar\mu})}{\Pi^{-1}_{\mu}(\omega)\Pi^{-1}_{1\mu}(\omega)
-\Sigma^{+}_{\bar\mu}(\omega)\Sigma^{-}_{\bar\mu}(\omega)}  \nl
&\qquad\qquad\quad
+\frac{i\big[\Pi^{-1}_{\mu}(\omega)-\Sigma^{-}_{\bar\mu}(\omega)\big]
n_{\bar\mu}}{\Pi^{-1}_{\mu}(\omega)\Pi^{-1}_{1\mu}(\omega)
-\Sigma^{+}_{\bar\mu}(\omega)\Sigma^{-}_{\bar\mu}(\omega)}
 \nl
&=
\frac{i(1-n_{\bar\mu})}
{\omega-\epsilon_\mu-\Sigma_{0\mu}
 +U\Sigma^+_{\bar\mu}(\omega-\epsilon_\mu-U-\Sigma_{0\mu}
 -\Sigma_{\bar\mu})^{-1}
 }
\nl&\quad
+\frac{ i n_{\bar\mu}}
{\omega-\epsilon_\mu-U-\Sigma_{0\mu}
 -U\Sigma^-_{\bar\mu}(\omega-\epsilon_\mu-\Sigma_{0\mu}
 -\Sigma_{\bar\mu})^{-1}
 }.
\end{align}
The frequency dependence of $\Sigma_{0\mu}$ and $\Sigma_{\bar\mu}$ is implied.
In this result, $n_{\bar\mu}=\bar\rho_{\bar\mu\bar\mu}+\bar\rho_{dd}$
and $1-n_{\bar\mu}=\bar\rho_{\mu\mu}+\bar\rho_{00}$.

\begin{figure}
\includegraphics[width=8cm]{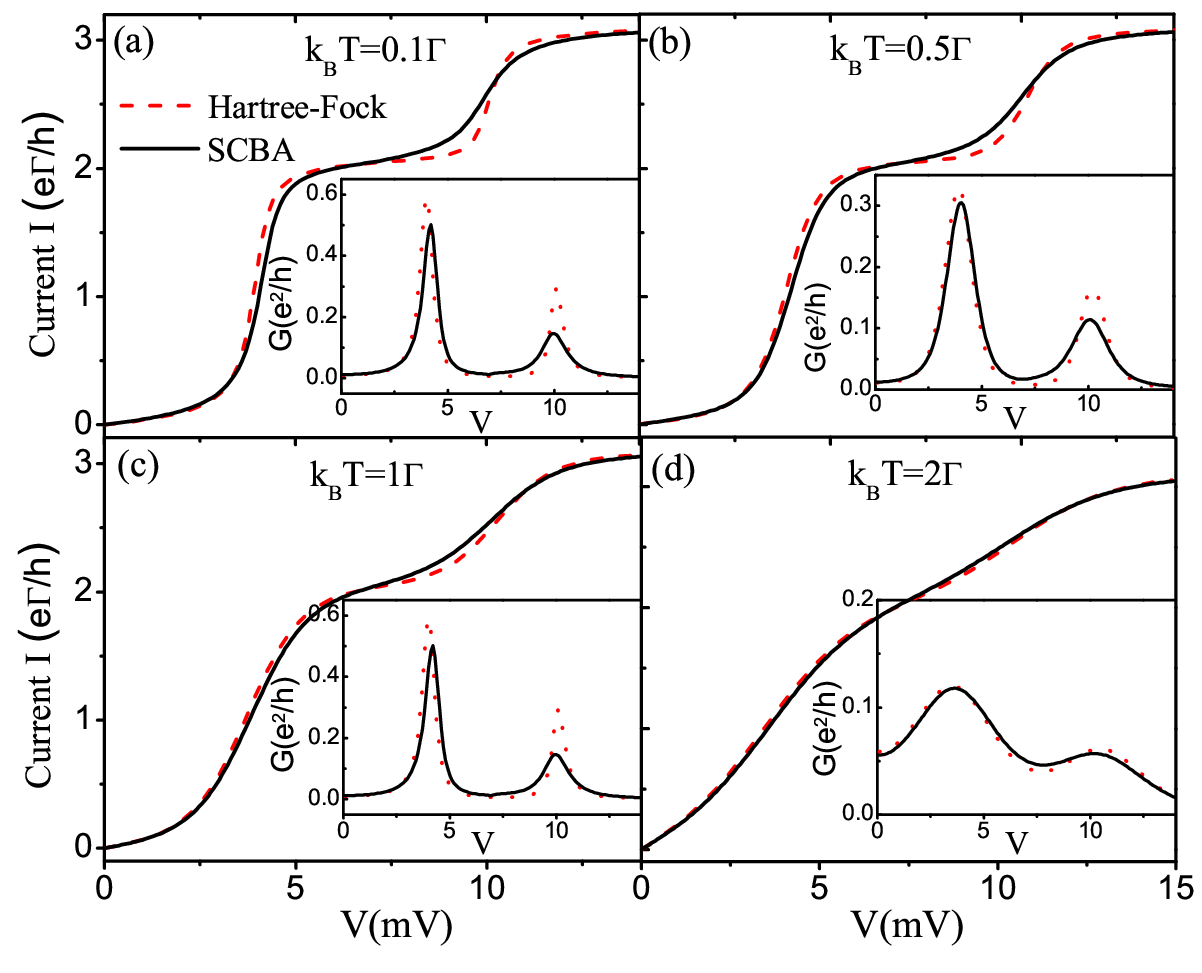}
\caption{(color online)
Coulomb staircase in the current-voltage curve.
Inset: the corresponding differential conductance.
Here, the results based on \Eq{Kondo} are plotted
against the  Hatree-Fock mean-field solution via \Eq{HF},
for the purpose of comparison.
Their difference gradually vanishes with the increase
of temperature, as shown from (a) to (d).
In the calculation, we consider $\Gamma_{\rm L}=\Gamma_{\rm R}=\Gamma/2$
and $\mu_{\rm L}=-\mu_{\rm R}=eV/2$. Taking the zero-bias Fermi level
as energy reference, we set $\epsilon_0=2$ meV, while assuming
$\Gamma=0.3$ meV and $U=3$ meV. }
\end{figure}

Equation (\ref{Kondo}) precisely coincides with the result from the nGF-EOM
formalism \cite{Hau96}.
 This solution, despite being certain overestimation
compared to other more sophisticated techniques \cite{Yan2012},
reveals qualitatively the remarkable nonequilibrium Kondo effect.
In Fig.\ 2 we display the Kondo peaks in the differential conductance.
This is a desirable result achieved in this work,
since the challenging Kondo effect is usually hard to be captured
by the conventional master equation methods,
including the second-order von Neumann approach
which goes also beyond the scope of the Born-Markov master equation \cite{Wac05+10},
and other approaches proposed more recently \cite{CS11,Leeu09,Galp09,Galp10}.

At high temperature, the Kondo physics, which is associated with
{\it coherent} forth-and-back tunneling,
is destroyed by the {\it incoherent} thermal process
between the dot and leads.
In this case, the $(\cdots)^{-1}$-terms
in the denominators in \Eq{Kondo} can be neglected,
resulting in
\be\label{HF}
 \varphi_{\mu\mu}(\omega) \approx \frac{i(1-n_{\bar\mu})}
 {\omega-\epsilon_\mu-\Sigma_{0\mu} }
 +  \frac{ i n_{\bar\mu}}
{\omega-\epsilon_\mu-U-\Sigma_{0\mu}  }.
\ee
This is the same result as that
derived from the equation of motion for two-particle
nGF formalism under a Hatree-Fock mean-field
approximation \cite{Hau96}.
Nevertheless, even this simplified result goes beyond the scope
of the second-order master equation, as evident
that \Eq{HF} does contain the {\it broadening effect}.
In Fig.\ 3 we plot the $I$-$V$ curves from both
\Eqs{Kondo} and (\ref{HF}) for comparison.
We would like to mention that,
in the Coulomb-blockade regime (the plateau stages),
the cotunneling contribution has automatically entered \Eq{Kondo},
in our unified treatment under the SCBA.      

\section{Summary}

We have proposed an efficient master equation approach to
quantum transport, by implementing a generalization
from the Born to self-consistent Born approximation.
We showed that the proposed scheme can give satisfactory results.
For instance, it can recover not only the exact
result of noninteracting transport under arbitrary voltages,
but also the nonequilibrium Kondo effect of interacting quantum dot.
This achievement goes beyond the scope of
the widely used master equation
under Born approximation, 
as well as other improved schemes
\cite{Wac05+10,CS11,Leeu09,Galp09,Galp10}.

As a final remark, compared to the nGF equation-of-motion 
formulated for the average of physical observables,
the master equation is for the (reduced) state. 
Thus it encodes more information and promises broader
applications beyond the steady-state current. 
In particular, the master equation approach is very 
useful for evaluating the shot noise
and full counting statistics on transport current.

\acknowledgments

Support from HNUEYT, the NNSF of China
(No.\ 10904029, 11274085, 91321106, \& 21033008),
Major State Basic Research Project of China
(No.\ 2011CB808502 \& 2012CB932704),
and the Hong Kong University
Grants Committee (AoE/P-04/08-2) and Research Grants
Council (No. 605012)
is gratefully acknowledged.

\appendix

\section{ Derivation of \Eq{tirhoj} }
\label{thapp-propa}

In this appendix, we present the derivation
of \Eq{tirhoj}, the equation-of-motion for
$\ti\rho_j(t)\equiv {\cal U}(t,\tau)[a^{\sigma}_\nu\rho(\tau)]$.
Similar to deriving the second-order master equation, we start
with a second-order expansion for the tunneling Hamiltonian $H_{SB}$,
and formally obtain the same equation as \Eq{cumm-expan}:
\be\label{tirhot0}
 \dot{\ti\rho}_{j}(t)=-i{\cal L}\ti\rho_{j}(t)-
\int^t_{0}\! d\tau{\rm Tr}_{B}\!
\big[{\cal L}'(t){\cal G}(t,\tau){\cal L}'(\tau)\ti\rho_{T}(\tau) \big].
\ee
The various superoperators in this equation have the same meaning
as in \Eq{cumm-expan}. Explicitly, the Liouvillian self-energy
superoperator can be reexpressed in Hilbert space via
\begin{align}
&\quad {\rm Tr}_{B}\big[{\cal L}'(t){\cal G}(t,\tau){\cal L}'(\tau)\ti\rho_{T}(\tau)\big]
\nl&=
  {\rm Tr}_{B}\big[H'(t)G(t,\tau)H'(\tau)\ti\rho_{T}(\tau)G^\dg(t,\tau)
\nl&\quad
 -G(t,\tau)H'(\tau)\ti\rho_{T}(\tau)G^\dg(t,\tau)H'(t)
\nl&\quad
-H'(t)G(t,\tau)\ti\rho_{T}(\tau)H'(\tau)G^\dg(t,\tau)
\nl&\quad
 +G(t,\tau)\ti\rho_{T}(\tau)H'(\tau)G^\dg(t,\tau)H'(t)
 \big]
\nl&
 \equiv  [I]-[II]-[III]+[IV].
\end{align}
Here, $H'(t)\equiv e^{iH_{B}t/\hbar}H_{SB}e^{-iH_{B}t/\hbar}$.
Applying the Born approximation,
$\ti\rho_{T}(\tau)\approx \rho^{\rm st}_{\rm B} \ti\rho_j(\tau)$,
with $\rho^{\rm st}_{\rm B}$ being the steady-state of the bare electrodes bath,
we further obtain
\bsube
\begin{align}
[I]&=
{\rm Tr}_B[H'(t)G(t,\tau)H'(\tau)\rho_{\rm B}
\ti\rho_j(\tau)G^\dg(t,\tau)]   \nl
&=
\sum_{\mu\nu} \Big\{C^{(+)}_{\mu\nu}(t-\tau)a_\mu {\cal G}(t,\tau)
\big[a^\dg_\nu\ti\rho_j(\tau)\big]
\nl&\quad\quad+ C^{(-)}_{\mu\nu}(t-\tau)a^\dg_\mu{\cal G}(t,\tau)
\big[a_\nu\ti\rho_j(\tau)\big]
\Big\},
\end{align}
\begin{align}
 [II] &=
{\rm Tr}_B[G(t,\tau)H'(\tau)\rho_{\rm B} \ti\rho_j(\tau)G^\dg(t,\tau)H'(t)]
\nl&= -
\sum_{\mu\nu} \Big\{C^{(+)}_{\mu\nu}(t-\tau) {\cal G}(t,\tau)
\big[a^\dg_\nu\ti\rho_j(\tau)\big]a_\mu
\nl&\quad\quad+
C^{(-)}_{\mu\nu}(t-\tau){\cal G}(t,\tau)\big[a_\nu\ti\rho_j(\tau)\big]a^\dg_\mu
\Big\},
\end{align}
\begin{align}
 [III] &=
{\rm Tr}_B[H'(t)G(t,\tau)\rho_{\rm B} \ti\rho_j(\tau)H'(\tau)G^\dg(t,\tau)]
\nl&= -
\sum_{\mu\nu} \Big\{C^{(-)\ast}_{\mu\nu}(t-\tau)a_\mu {\cal G}(t,\tau)
\big[\ti\rho_j(\tau)a^\dg_\nu\big]
\nl&\quad\quad+
C^{(+)\ast}_{\mu\nu}(t-\tau)a^\dg_\mu{\cal G}(t,\tau)\big[\ti\rho_j(\tau)a_\nu\big]
\Big\},
\end{align}
\begin{align}
 [IV]&=
{\rm Tr}_B[G(t,\tau)\rho_{\rm B} \ti\rho_j(\tau)H'(\tau)G^\dg(t,\tau)H'(t)]
\nl&=
\sum_{\mu\nu} \Big\{C^{(-)\ast}_{\mu\nu}(t-\tau) {\cal G}(t,\tau)
\big[\ti\rho_j(\tau)a^\dg_\nu\big]a_\mu
\nl&\quad\quad+
C^{(+)\ast}_{\mu\nu}(t-\tau){\cal G}(t,\tau)
\big[\ti\rho_j(\tau)a_\nu\big]a^\dg_\mu
\Big\}.
\end{align}
\esube
Substituting these results into \Eq{tirhot0},
a more compact notation leads to \Eq{tirhoj}.

It is of crucial importance to note that, in the above
$[II]$ and $[III]$, extra minus sign appears
when we rearrange the bath operators to the two sides of $\rho_B$.
We explain this issue in more detail as follows.
Consider, for instance,
${\rm Tr}_B \{[a^{\dg}_{\nu}F_{\alpha\nu}(\tau)]
[\rho_B \tilde{\rho}_j(\tau)]
[F^{\dg}_{\alpha\mu}(t) a_{\mu}]  \}$.
In order to utilize the cyclic property under ${\rm Tr}_B[\cdots]$,
i.e., ${\rm Tr}_B [F_{\alpha\nu}(\tau)\rho_B F^{\dg}_{\alpha\mu}(t) ]
= C^{(+)}_{\alpha\mu\nu}(t-\tau)$,
we have to move $F^{\dg}_{\alpha\mu}(t)$, crossing $\tilde{\rho}_j(\tau)$,
to the right side of $\rho_B$.
Since $\tilde{\rho}_j(\tau)$ contains a Fermi operator,
$a_j\equiv a^{\sigma}_{\nu}$, the aforementioned move
of $F^{\dg}_{\alpha\mu}(t)$ will cause an additional minus sign,
according to the Fermi-Dirac anticommutative relation that
$a_jF^{\dg}_{\alpha\mu}=-F^{\dg}_{\alpha\mu}a_j$.
As a consequence, this type of extra minus sign
alters the {\it commutators} in \Eq{rhot-second}
to the {\it anti-commutators} in \Eq{tirhoj}.


\section{ Derivation of \Eq{varphiw0} }
\label{thapp-varhpiw0}

Starting with the definition $\varphi_{i\mu\nu}(\omega)
={\rm Tr} \big[a_\mu\ti\rho_{i\nu}(\omega)\big]$
and \Eq{tirhow}, we have
\begin{align}\label{varphiapp}
&-i\omega \varphi_{i\mu\nu}(\omega)-\varphi_{i\mu\nu}(0)
\nl&
 = -i {\rm Tr} \big\{[a_\mu,H]\ti\rho_{i\nu}(\omega)\big\}
 - {\rm Tr}\big[a_\mu\Sigma(\omega)\ti\rho_{i\nu}(\omega)\big].
\end{align}
For noninteracting system, we process the first term in the right-hand-side
of \Eq{varphiapp}: ${\rm Tr} \big\{[a_\mu,H]\ti\rho_{i\nu}(\omega)\big\}
=\sum_m h_{\mu m}{\rm Tr} \big[a_m \ti\rho_{i\nu}(\omega)\big]
 =\sum_m h_{\mu m} \varphi_{i m\nu}$, i.e.,
\be\label{varphiapp1}
{\rm Tr} \big\{[a_\mu,H]\ti\rho_{i\nu}(\omega)\big\}
=[\bm h \bm \varphi_i]_{\mu\nu}.
\ee
As in the main text, we introduce the boldface matrix notation for brevity.
Further, we process the second term in \Eq{varphiapp}:
\begin{align}
\Sigma(\omega)\ti\rho_{i\nu}(\omega)
&=\sum_{mn}\Big[C^{(+)}_{nm}(\omega-{\cal L})
\big\{a_n, a^\dg_m \ti\rho_{i\nu}(\omega)\big\}
\nl&\quad\quad
 +C^{(-)}_{nm}(\omega-{\cal L})
\big\{a^\dg_n, a_m \ti\rho_{i\nu}(\omega)\big\}
\nl&\quad\quad
  +C^{(+)\ast}_{nm}({\cal L}-\omega)
\big\{a^\dg_n, \ti\rho_{i\nu}(\omega)a_m \big\}
\nl&\quad\quad
  +C^{(-)\ast}_{nm}(\omega-{\cal L})
\big\{a_n, \ti\rho_{i\nu}(\omega) a^\dg_m \big\}\Big].
\end{align}
Using the anti-commutative relation of fermions,
$\{a_\mu,a^\dg_\nu\}=\delta_{\mu\nu}$,
and the cyclic invariance property under trace, which leads to
${\rm Tr}\big[a_\mu\{a^\dg_n,a_m \ti\rho_{i\nu}(\omega)\}\big]
=\delta_{n\mu} \varphi_{i m\nu}(\omega)$ and
${\rm Tr}\big[a_\mu\{a_n,a^\dg_m \ti\rho_{i\nu}(\omega)\}\big]
=0$,
we obtain
\begin{align}\label{varphiapp2}
{\rm Tr}\big[a_\mu\Sigma(\omega)\ti\rho_{i\nu}(\omega)\big]
&=i\sum_{m}\Sigma_{0\mu m}(\omega)\varphi_{i m \nu}
\nl&
 =i\big[\bm \Sigma_{0}(\omega)\bm \varphi_{i}\big]_{\mu \nu},
\end{align}
where $\Sigma_{0\mu\nu}(\omega)=
-i\big[C^{(-)}_{\mu \nu}(\omega)+C^{(+)\ast}_{\mu \nu}(-\omega)\big]$.
Then, inserting \Eqs{varphiapp1} and (\ref{varphiapp2})
into \Eq{varphiapp}, we arrive at \Eq{varphiw0}.

\section{Derivation of \Eqs{calUw}--(\ref{lakaw})}
\label{thapp-III}

The most direct way to get the solution of \Eq{calUw}
is to express all the superoperators, such as ${\cal L}$
and $\Sigma(\omega)$, in Liouvillian space
which is expanded by $\{|mn\ra\ra\equiv|m\ra\la n|\}$
with $m, n=0,\up,\down, d$.
Using MATHEMATICA, one can analytically inverse the matrix \Eq{Uomega},
then obtain ${\cal U}(\omega)$ in Liouvillian space in terms of a
$16\times 16$ matrix form and the solution of \Eq{calUw}.
However, this type of solution is too lengthy to be presented here.
---This solving scheme is more appropriate for numerical calculations.

For the specific problem considered here, we prefer
a more compact way to obtain \Eq{calUw} as follows.
Based on \Eq{tirhow}, we plan to solve for $\ti\rho_{1\mu}(\omega)
\equiv {\cal U}(\omega)[a^\dg_\mu \bar\rho]$
and $\ti\rho_{2\mu}(-\omega)
\equiv {\cal U}(-\omega)[a_\mu \bar\rho]$,
instead of ${\cal U}(\pm\omega)$
since most of its matrix elements are zero.
Here, as an example, we outline the derivation for
$\ti\rho_{1\mu}(\omega)$, under the initial condition of
$\ti\rho_{1\mu}(0)=a^{\dg}_\mu\bar\rho$.
For the Anderson impurity model,
spin conservation would make the steady-state density matrix
diagonalized in the specified state basis. 
Accordingly, we have
$\ti\rho_{1\mu}(0)=a^{\dg}_\mu\bar\rho=\bar\rho_{00}|\mu\ra\la 0|
+(-1)^{\mu}\bar\rho_{\bar\mu\bar\mu}|d\ra\la \bar \mu|$
and obtain [c.f. the first identity in \Eq{calUw}]:
\be\label{tirho1mu}
\ti\rho_{1\mu}(\omega) = \lambda^+_\mu(\omega)|\mu\ra\la 0|
+(-1)^{\mu}\kappa^+_\mu(\omega)|d\ra\la \bar \mu|
\ee
where $\lambda^+_\mu(\omega)=\la\mu|\ti\rho_{1\mu}(\omega)|0\ra$
and
$\kappa^+_\mu(\omega)=\la d|\ti\rho_{1\mu}(\omega)|\bar\mu\ra$,
with initial conditions of
$\lambda^+_\mu(0)=\bar\rho_{00}$
and $\kappa^+_\mu(0)=\bar\rho_{\bar\mu\bar\mu}$, respectively.
In constructing \Eq{tirho1mu}, we have implemented
the following considerations.
First, the basis states are eigenstates of $H_S$.
Second, the self-energy term of \Eq{tirhow}
does not mix the diagonal and off-diagonal matrix elements.
Therefore, the structure of $\ti\rho_{1\mu}(\omega)$,
i.e., the form of having nonzero matrix elements,
is identical to that of $\ti\rho_{1\mu}(0)$.

Inserting \Eq{tirho1mu} into \Eq{tirhow}, we obtain
\be\label{eq_C2}
\begin{split}
-i\Pi^{-1}_\mu(\omega)\lambda^+_\mu(\omega)
&=\lambda^+_\mu(0)+i\Sigma^-_{\bar\mu}(\omega) \kappa^+_\mu(\omega),
\\
-i\Pi^{-1}_{1\mu}(\omega)\lambda^+_\mu(\omega)
&=\kappa^+_\mu(0)+i\Sigma^+_{\bar\mu}(\omega) \lambda^+_\mu(\omega).
\end{split}
\ee
Then, $\lambda^+_\mu(\omega)$ and $\kappa^+_\mu(\omega)$
can be easily carried out.
Using the same method outlined above,
one can obtain $\lambda^-_\mu(\omega)$ and $\kappa^-_\mu(\omega)$,
and solve for $\ti\rho_{2\mu}(\omega)$ under the initial condition
$\lambda^+_\mu(0)=\bar\rho_{\mu\mu}$ and $\kappa^+_\mu(0)=\bar\rho_{dd}$.
Finally, we mention that the solution of $\ti\rho_{1\mu}(\omega)$
and $\ti\rho_{2\mu}(\omega)$, quite straightforwardly, gives the result
of \Eq{Kondo} via
$\varphi_{\mu\mu}(\omega)={\rm Tr} \big[a_\mu\ti\rho_{1\mu}(\omega)\big]
+{\rm Tr} \big[a_\mu\ti\rho_{2\mu}(\omega)\big]$.

\section{Relation with the hierarchical master equation theory}
\label{app_HEOM}

The SCBA-ME is constructed by an insight from the Feynman's
diagrammatic technique on Keldysh contour.
In this appendix, we build its connection to
the HEOM approach developed
recently on the basis of the Feynman-Vernon influence
functional theory for quantum open systems \cite{Yan080911}.
For this purpose, we introduce first the notation used
in Ref.\ \cite{Yan080911}:
\bsube\label{eq_D1}
\begin{align}
\rho^{(\sigma)}_{\alpha\mu}(t)
& =
-i\left[{\cal A}^{(\sigma)}_{\alpha\mu\rho}(t)
-{\cal A}^{(\bar\sigma)\dg}_{\alpha\mu\rho}(t)\right],
\\
\phi^{(\sigma)}_{\alpha\mu}(\omega,t)
& =
-i\left[\ti {\cal A}^{(\sigma)}_{\alpha\mu\rho}(\omega,t)
-\ti {\cal A}^{(\bar\sigma)\dg}_{\alpha\mu\rho}(\omega,t)\right],
\label{phiA}
\end{align}
\esube
where ${\cal A}^{(\sigma)}_{\alpha\mu\rho}(t)
=\int \frac{d\omega}{2\pi}\ti {\cal A}^{(\sigma)}_{\alpha\mu\rho}(\omega,t)$,
with [c.f.\ \Eq{Arho-SCBA}]
\be\label{eq_D2}
 \ti {\cal A}^{(\sigma)}_{\alpha\mu\rho}(\omega,t)
= \!
 \sum_\nu \!\!\int^t_0 \!\! d\tau \Gamma^{(\sigma)}_{\alpha\mu\nu}
 (\omega)e^{i\sigma\omega (t-\tau)}
 \big\{{\cal U}(t,\tau)[a^{\sigma}_\nu\rho(\tau)]\big\}.
\ee
Compared to Ref.\ \cite{Yan080911}, we find
$\rho^{(\sigma)}_{\alpha\mu}(t)= \int \frac{d\omega}{2\pi}
\phi^{(\sigma)}_{\alpha\mu}(\omega,t)$, which is nothing but
the first-tier auxiliary density operator introduced there.

 Now consider the quantity in the curry brackets in \Eq{eq_D2}.
Formally, from \Eq{tirhoj} we have
\be\label{eq_D3}
\partial_t{\cal U}(t,t_0)=-i{\cal L}{\cal U}(t,t_0)
-i\sum_{\alpha,\mu,\sigma}\int \frac{d\omega}{2\pi}
 \big\{ a^{\bar\sigma}_\mu,{\cal U}^{(\sigma)}_{\alpha\mu}(\omega,t)\big\}, 
\ee
where ${\cal U}^{(\sigma)}_{\alpha\mu}(\omega,t)\equiv
{\cal U}^{(\sigma)}_{\alpha\mu}(\omega,t,t_0)$ is implied and satisfies
\be\label{eq_D4}
 \partial_t{\cal U}^{(\sigma)}_{\alpha\mu}(\omega,t)
 =-i({\cal L}-\sigma\omega){\cal U}^{(\sigma)}_{\alpha\mu}(\omega,t)
-i{\cal C}^{(\sigma)}_{\alpha\mu+}(\omega){\cal U}(t,t_0).
\ee
Here, ${\cal C}^{(\sigma)}_{\alpha\mu+}(\omega)$ is a superoperator, defined via
\be\label{eq_D5}
{\cal C}^{(\sigma)}_{\alpha\mu\pm}(\omega)\hat O=
  \sum_\nu\big[\Gamma^{\sigma}_{\alpha\mu\nu}(\omega)
   a^\sigma_\nu\hat O
   \pm\Gamma^{\bar\sigma}_{\alpha\nu\mu}(\omega)\hat O
   a^\sigma_\nu\big].
\ee
With these identifications, we can now recast \Eq{QME-SCBA} as
\bsube\label{HEOM2}
\begin{align}
 &\dot\rho(t)
 =-i{\cal L}\rho(t)-i\sum_{\alpha\mu\sigma}\int \frac{d\omega}{2\pi}
 \big[ a^{\bar\sigma}_\mu,\phi^{(\sigma)}_{\alpha\mu}(\omega,t)\big],
\label{HEOM2a} \\
 &\dot\phi^{(\sigma)}_{\alpha\mu}(\omega,t)
=-i({\cal L}-\sigma\omega)\phi^{(\sigma)}_{\alpha\mu}(\omega,t)
 -i{\cal C}^{(\sigma)}_{\alpha\mu-}(\omega)\rho(t)
\nl&\qquad\qquad\quad
 -i\!\!\sum_{\alpha'\mu'\sigma'}\!\!\int d\omega'
 \big[ a^{\bar\sigma'}_{\mu'},\phi^{(\sigma\sigma')}_{\alpha\mu\alpha'\mu'}
 (\omega,\omega',t)\big],
\label{HEOM2b} \\
 &\dot\phi^{(\sigma\sigma')}_{\alpha\mu\alpha'\mu'}(\omega,\omega',t)
  =-i({\cal L}-\sigma\omega-\sigma'\omega')
\phi^{(\sigma\sigma')}_{\alpha\mu\alpha'\mu'}(\omega,\omega',t)
\nl&\qquad\qquad\qquad\qquad
 -i{\cal C}^{(\sigma')}_{\alpha'\mu'+}(\omega')
 \phi^{(\sigma)}_{\alpha\mu}(\omega,t).
\label{HEOM2c}
\end{align}
\esube
This form of SCBA-ME resembles the second-tier-level HEOM
\cite{Yan080911}, but with one difference:
\Eq{HEOM2c} does not have
the term of $-i{\cal C}^{(\sigma)}_{\alpha\mu+}(\omega)
\phi^{(\sigma')}_{\alpha'\mu'}(\omega',t)$.
As $\phi^{(\sigma)}_{\alpha\mu}(\omega,t)$ of \Eq{HEOM2b}
is concerned, this neglected term represents the
nonlocal $\{\omega';\sigma'\alpha'\mu'\}$-contributions.
In fact the transport current
and the effective self-energy are dictated explicitly only with
$\{\phi^{(\sigma)}_{\alpha\mu}(\omega,t)\}$.
Apparently, the inclusion of those nonlocal contributions
will significantly compromise the numerical efficiency
in evaluating the desired
$\int d\omega'\,\phi^{(\sigma\sigma')}_{\alpha\mu\alpha'\mu'}(\omega,\omega',t)$
for $\phi^{(\sigma)}_{\alpha\mu}(\omega,t)$ of \Eq{HEOM2b}.
As highlighted in \Sec{thsec2B}, SCBA-ME
exploits the so-called random-phase approximation,
which assumes those nonlocal contributions
be collectively negligible.
With this ansatz \Eq{HEOM2b} is effectively
a single-frequency ($\omega$) task in evaluation.
More importantly, this ansatz is found to be truly valid,
at least for all the cases of study in this work.

 As demonstrated in the main text of this work,
SCBA-ME that is equivalent to \Eq{HEOM2} does give satisfactory results.
This is in good agreement with other SCBA scenarios in physics.
On the other hand, in applying HEOM  \cite{Yan080911,Yan2012}, it was often found
the numerical satisfactory results at the second-tier
level of truncation.
This consistency convinces us that the SCBA-ME
should be a valuable quantum transport approach,
with a compact form for practical manipulation
and a reasonable accuracy.

\section{ Lorentzian reservoir spectral density }

In most cases for a system coupled to a continuum, the Lorentzian type
spectral density function is more reasonable than a {\it constant} one.
In quantum transport, for the coupling to the leads, we therefore assume
$\Gamma_{\alpha\mu \nu}(\omega)=2\pi\sum_k t_{\alpha\mu k} t^\ast_{\alpha\nu k}
\delta(\omega-\epsilon_{\alpha k})$ as
\cite{Yan080911}
\be\label{Gammaw}
\Gamma_{\alpha\mu \nu}(\omega)=
\frac{\Gamma_{\alpha\mu \nu} W^2_\alpha}{(\omega-\mu_\alpha)^2+W^2_\alpha} .
\ee
Strictly speaking, this corresponds to a half-occupied band
for each lead, which centers the Lorentzian function
at the Fermi level of the lead.
However, since the Fermi level can locate anywhere,
depending on the electron density for instance,
it is better to understand \Eq{Gammaw}
as for model studies, which has the advantage
of modeling a finite bandwidth and leading to some
compact (analytic) expressions.
In \Eq{Gammaw}, $W_\alpha$ characterizes the bandwidth of the $\alpha$-th lead.
Quite naturally, the constant coupling rate can be recovered
by assuming $W_\alpha\rightarrow\infty$,
yielding $\Gamma_{\alpha\mu \nu}(\omega)=\Gamma_{\alpha\mu \nu}$.   

Corresponding to the above Lorentzian spectral density,
the correlation function of \Eq{FDT2} can be expressed as
\begin{align}
C^{(\pm)}_{\alpha\mu \nu}(\omega)
&=\frac{1}{2}\left[\Gamma^{(\pm)}_{\alpha\mu \nu}(\mp \omega)
 +i\Lambda^{(\pm)}_{\alpha\mu \nu}(\mp \omega)\right].
\end{align}
The second quantity is related to the first one through
the well-known dispersive relation as
\begin{align}
\Lambda^{(\pm)}_{\alpha\mu \nu}(\omega)
&={\cal P}\int^\infty_{-\infty}\frac{d\omega'}{2\pi}
\frac{1}{\omega\pm\omega'}\Gamma^{(\pm)}_{\alpha\mu \nu}(\omega)
\nl&=\frac{\Gamma_{\alpha\mu\nu}}{\pi}
\Bigg\{{\rm Re}\left[\Psi\left(\frac{1}{2}
+i\frac{\beta(\omega-\mu_\alpha)}{2\pi}\right)\right]
\nla
-\Psi\left(\frac{1}{2}+\frac{\beta W_\alpha}{2\pi}\right)
\mp\pi\frac{\omega-\mu_\alpha}{W_\alpha}\Bigg\},
\end{align}
where ${\cal P}$ denotes the principle value of the integral,
and $\Psi(x)$ is the digamma function.


\end{document}